\documentclass[pra,aps,twocolumn,showpacs]{revtex4}

\usepackage{graphicx}
\usepackage{amssymb}
\usepackage{graphicx}
\usepackage{bm}
\usepackage{amsmath}

\begin{document}

\title{Trapping cold atoms near carbon nanotubes:\\
thermal spin flips and Casimir--Polder potential}

\author{R.~Fermani}
\email{rachele.fermani@imperial.ac.uk}
\author{S.~Scheel}
\author{P.L.~Knight}

\affiliation{Quantum Optics and Laser Science, Blackett
Laboratory, Imperial College London, Prince Consort Road, London
SW7 2AZ, United Kingdom}

\date{\today}

\begin{abstract}
We investigate the possibility to trap ultracold
$^{87}\mathrm{Rb}$ atoms near the outside of a metallic carbon
nanotube (CN) which we imagine to use as a miniaturized
current-carrying wire. We calculate atomic spin flip lifetimes and
compare the strength of the Casimir--Polder potential with the
magnetic trapping potential. Our analysis indicates that the
Casimir--Polder force is the dominant loss mechanism and we
compute the minimum distance to the carbon nanotube at which the
atoms can be trapped to be larger than $100$ nm.
\end{abstract}

\pacs{ 34.50.Dy, 42.50.Ct, 78.67.Ch, 03.75.Be}

 \maketitle
\section{Introduction}

Advances in magnetic trapping of clouds of ultracold atoms and
Bose-Einstein condensates have received considerable research
attention \cite{HINDSHUGES,HANSEL,FOLMAN,REICHEL02}. The control
and manipulation of atomic clouds is of fundamental importance in
the investigation of the basic physical properties of atom-surface
interaction \cite{HENKEL00,JONES,FERMANI}, as well as in quantum
information processing (QIP) \cite{SCHMIEDMAYER,JAKSCH1999}. As
trapped cold atoms appear to be very sensitive to magnetic-field
variations, they represent a powerful tool in magnetic field
imaging as well as in gaining insight into atom-surface coupling
phenomena \cite{SCHMIEDMAYER_NATURE,SCHMIEDMAYER_06}.  The
challenge is to keep the atoms as close as possible to the
substrate material to map the magnetic and electric fields in the
vicinity of the surface. The combination of quantum state control
with the development of ever-smaller magnetic traps is an
essential element in the implementation of integrated quantum
devices for fundamental research, quantum information processing
and precision measurement.

Along with the push towards miniaturization evolved the idea of
devising even smaller structures based on carbon nanotubes (CNs)
\cite{PEANO}. Carbon nanotubes are carbon mono-layers rolled-up
into cylinders of a few nanometers' diameter
\cite{IJIMA,BENEDICT}. They have been widely investigated
theoretically and experimentally as they play a key role in
miniaturized electronic, mechanical, electromechanical and
scanning-probe devices. For their potential use as miniaturized
current-carrying wires it is important to realize that the desired
close proximity of a neutral atom and the carbon atoms that make
up the nanotube can vastly enhance the influence of dispersion
forces, and we address this point in the paper.

It is well known that an atom held in a magnetic trap near an
absorbing dielectric surface will be subject to thermally-induced
spin flip transitions whose origin lies in fluctuating magnetic
fields which can be attributed to resistive noise in the substrate
\cite{Henkel99,SCHEEL}. In accordance with the
fluctuation-dissipation theorem, dissipation processes associated
with a finite  conductivity give rise to electromagnetic-field
fluctuations. These fluctuations can be strong enough to drive
spin transitions that lead to trapping losses
\cite{EXPATOMCHIP,Cornell,Vuletic}.

In addition, an atom placed near a dielectric body will experience
a dispersion force due to the presence of the dielectric material
--- the Casimir--Polder force \cite{PITAV,MILONNI,Blagov,Obrecht}. The
potential generating this force adds to the magnetic trapping
potential and may cause the trap to become unstable at small
distances.

With this in mind, CNs seem to represent rather attractive
structures for designing miniaturized magnetic traps. This is on
one hand due to the fact that they consist of a very small amount
of dielectric matter which means that unwanted dispersion forces
such as the Casimir--Polder force are minimized. On the other
hand, they also possess extremely homogeneous surfaces and are
thus less likely to induce inhomogeneities in the potential
surface of the trap.

This paper is organized as follows. In Sec.~\ref{sec:potential} we
introduce the basic concepts of magnetic trapping of neutral
atoms. In Sec.~\ref{sec:lifetime} the spin flip lifetime is
calculated and compared to the tunneling lifetime resulting from
the combination of the magnetic trapping potential and the
Casimir--Polder potential. Both lifetimes are given in terms of
the dyadic Green tensor. In order not to interrupt the flow of
arguments and results we will present technical details about the
conductivity and the Green tensor for a single wall CN in the
Appendix. Some conclusions are drawn in Sec.~\ref{sec:conclusion}.


\section{Magnetic traps for neutral atoms}
\label{sec:potential}

In this section we briefly review magnetic trapping using
straightforward tools of electromagnetism. An atom with a magnetic
dipole moment $\bm{\mu}$ placed in a magnetic field $\mathbf{B}_{T}$
experiences an interaction potential
\begin{equation}\label{eq:potential}
V= - \bm{\mu} \cdot \mathbf{B}_{T} .
\end{equation}
Assuming that a low-field seeking atom is in the state $|I J F
m_{F}\rangle$, Eq.~(\ref{eq:potential}) corresponds to a Zeeman
energy $V= g_{F} \mu_{B} m_{F} B$ that depends on the quantum
number $m_{F}$ and on the magnitude of the field
$B=|\mathbf{B}_{T}|$ while it is independent of the field's
direction, with $g_{F}$ the Land\'{e} factor and $\mu_{B}$
the Bohr magneton. Let us consider a current $I$ flowing through a
wire along the $z$ direction that generates a circular magnetic
field $\mathbf{B}$.

Applying a homogeneous bias magnetic field $\mathbf{B}_{b}$
pointing in a direction orthogonal to the wire, a line of
vanishing magnetic field parallel to the wire is created which is
located a distance $y_0=\mu_{0}I/(2\pi B_b)$ away from the wire.
The total field is given by $\mathbf{B}_T=\mathbf{B}+\mathbf{B}_b$
whose gradient $B'(r)= -\mu_{0}I/(2\pi r^2)$, at the position of
the field minimum $y_0$, can be written as
$B'(y_0)=b'=-2B_b^2\pi/(\mu_0 I)$. As $B'(y_0)$ is independent of
position, the superposition of the magnetic field created by the
wire and the homogeneous bias field creates a two-dimensional
quadrupole-type trap \cite{REICHEL}. In such a trap, the trapping
potential can be approximated by a linear function of the magnetic
field gradient $b'$. The magnetic field can be expressed as
$\mathbf{B}=b'x\mathbf{e}_{x}-b'y\mathbf{e}_{y}$ and its modulus
is $|\mathbf{B}|=b'(x^{2}+y^{2})^{1/2}=b'r$ with $r$ denoting the
distance from the trap centre.

The modulus of the $y$-component of $\mathbf{B}_{T}$ vanishes at
the trap center. In order to prevent Majorana transitions to
non-trapped magnetic levels, a further offset field $\mathbf{B}_o$
parallel to the wire is applied with $|\mathbf{B}_o|\ll
|\mathbf{B}|$. The magnitude of the field near the center of the
trap is then given by \cite{FOOT}
\begin{equation}
|\mathbf{B}| = \left[ {B_{o}^{2}+(b' r)^{2}} \right]^{1/2}
\approx B_{o}+ \frac{b'^{2}r^{2}}{2 B_{o}},
\end{equation}
where the approximation holds for $b' r \ll B_{o}$. The presence
of the $\mathbf{B}_o$ field changes the shape of the potential
near the trap centre from being linear to harmonic. The interaction
potential that follows from Eq.~(\ref{eq:potential}) is then, to a
good approximation, a harmonic potential of the form
$V=V_0+\frac{1}{2} M \omega_r^2 r^2$ where $M$ denotes the mass of the
atom. The trap oscillation frequency $\omega_r$ is given by
\begin{equation}
\label{eq:trapfrequency}
\omega_{r}=\sqrt{\frac{g_{F} \mu_B m_F}{MB_{0}}}
\frac{\mu_{0}I}{2 \pi y_{0}^{2}} \,.
\end{equation}
Moreover, $\mathbf{B}_o$ controls both splitting of the magnetic
sublevels by a frequency $f_0=\frac{1}{2} \mu_B B_o/h$ at the trap
center as well as the stability of the resonance associated with the
magnetic guide \cite{Bill}.

\section{Trapping lifetimes}
\label{sec:lifetime}

In this section, we investigate the two main limitations to the
trapping lifetime, thermally-induced spin flip transitions and the
Casimir--Polder potential. In the following calculations, we
take the current through the single-wall CN to be equal to
$I=20\,\mu$A which seems to be the largest current that can be
sustained before saturation effects become important
\cite{CNcurrent}. The physical properties of a CN are determined by
the way in which the graphite sheet is rolled. The winding angle with
respect to the hexagonal carbon lattice is usually described by two
integer numbers $(a,b)$ \cite{HAMADA,TASAKI,LIN-SHUNG,JIANG}. When
$2a+b=3n$, where $n$ is again an integer, a CN shows metallic
behaviour, otherwise it is semi-conducting.

The axial conductivity $\sigma_{zz}(\omega)$ and the resulting
dielectric permittivity $\varepsilon (\omega)$ of a $(9,0)$ carbon
nanotube are calculated in Appendix~\ref{sec:epsilon}. At a
frequency $f_0=70$~kHz, chosen as to correspond to an offset field
$B_{o}=10^{-5}\,\mbox{T}=100\,\textrm{m}\mbox{G}$, we obtain
$\sigma_{zz}(\omega_{0})=1.19\cdot
10^{9}+11.5i\;(\Omega\mathrm{m})^{-1}$ and $\varepsilon
(\omega_{0})\simeq 3\cdot 10^{14}i$. Hence, a $(9,0)$ carbon
nanotube can indeed be considered as a metallic cylinder.

Thermal fluctuations generate noise currents that lead to fluctuating
fields near the body surface. We expect the noise due to these
fluctuating fields to be much reduced in a CN compared to a dielectric
bulk material due to the very small amount of matter
involved. Nevertheless, thermal spin flips and the Casimir--Polder
force cannot be neglected and needs investigation and a comparison of
their effect.
Both mechanisms originate from the fluctuations of the electromagnetic
field in the substrate. In particular, the spin-flip transitions are
caused by the magnetic-field fluctuations while the Casimir-Polder
force arises from both electric and magnetic field fluctuations, the
latter usually being negligible.

In order to describe the two phenomena we utilize the
quantization scheme of the electromagnetic field in the presence
of dispersing and absorbing bodies \cite{SCHEEL/98,PERINA}.
As this theory is a macroscopic theory whose central quantities
are linear susceptibilities, carbon nanotubes are probably at the
limit to what we can actually describe with it. However, when
viewed from distances that are several multiples of the bond
lengths, the CN can be thought of as a homogeneous object so that
the detailed structure from the surface cannot be resolved and QED
in dielectrics can be safely used. This also assumes that the CN
contains no impurities and shows no pitch alterations.

\subsection{Spin flip lifetime}
\label{sec:spinflip}

If an atom is held sufficiently close to the CN surface it will
experience quantum fluctuations of the electromagnetic field. At
the center of the trap, the atom feels a constant magnetic field
$B_o$. The atomic magnetic sublevels are thus split by the Zeeman
interaction and only a subset of these levels will experience an
attractive force (low-field seeking states). An $^{87}\mathrm{Rb}$
atom can be trapped in the hyperfine state
$|F,m_F\rangle=|2,2\rangle$ is trapped, but only for sufficiently
tight magnetic traps also in the state
$|F,m_F\rangle=|2,1\rangle$. Transitions to lower magnetic
sublevels allow the atom to escape. In the following, we disregard
all the lower-lying states and treat the atom in the two-level
approximation as the transition is the rate limiting step.

The lifetime of an atom due to spin flip transitions is given
by the inverse of the spin flip rate \cite{SCHEEL}
\begin{eqnarray}
\label{eq:gamma} \lefteqn{ \label{A1} \Gamma= \frac{2(\mu_B g_S)^2
}{c^2\varepsilon_0 \hbar} \,\langle f|\hat{S}_q| i\rangle\,
\langle i|\hat{S}_k| f\rangle } \nonumber \\ && \times \mathrm{Im}
\left[ \overrightarrow{\bm{\nabla}}\times
\bm{G}(\mathbf{r},\mathbf{r},\omega_0) \times
\overleftarrow{\bm{\nabla}}\right]_{qk}
\end{eqnarray}
where $\mu_{B}$ is the Bohr magneton, $\hat{S}_k$ is the $k$th vector
component of the electronic spin operator, and $g_{S}\approx 2$ the
electron's $g$ factor. Spin flips occur between the initial state
$|i\rangle$ and the final state $|f\rangle$, the position $\mathbf{r}$
of the atom is taken to be the centre of the trap.

The spin flip rate in Eq.~(\ref{eq:gamma}) is given in terms of
the dyadic Green tensor $\bm{G}(\mathbf{r},\mathbf{r},\omega)$
which contains the physical and geometrical information about the
nanotube. We assume the CN to be in thermal equilibrium
with the environment at a temperature $T$. The total spin flip rate is
then given by
$\Gamma_{\mathrm{tot}}=\Gamma(\bar{n}_{\mathrm{th}}+1)$ where
$\bar{n}_{\mathrm{th}}$ is the mean thermal occupation number
$\bar{n}_{\mathrm{th}}=(e^{\hbar \omega_{0}/k_BT}-1)^{-1}$, with
$k_B$ denoting Boltzmann's constant.

\begin{figure}[ht]
\includegraphics[width=8.4cm]{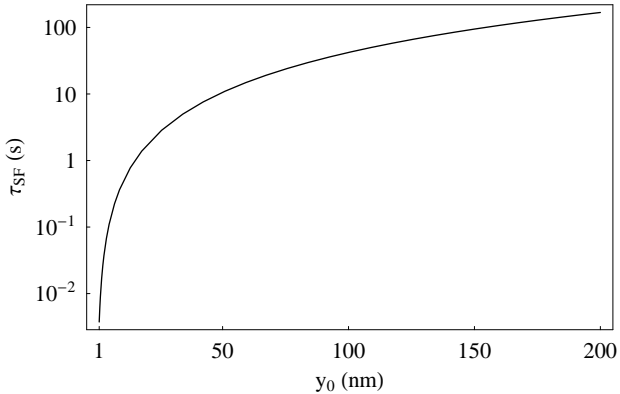}
\caption{\label{fig:lifetimeSF} Spin flip lifetime of a rubidium
atom near a $(9,0)$ carbon nanotube with radius
$R_{CN}=3.52$~{\AA}. The trapping distance $y_0$ is varied between
$1$ and $200$~nm. The other parameters are: $f_0=70$~kHz and
$T=380$~K.}
\end{figure}
In Fig.~\ref{fig:lifetimeSF} we show the calculated spin flip
lifetime $\tau_{SF}=1/\Gamma_{\mathrm{tot}}$ as a function of the
trapping distance $y_{0}$ from the surface of a $(9,0)$ CN for a
temperature $T=380$~K,
corresponding to a thermal excitation energy of
$k_BT=5.2\,10^{-21}$ J ($\simeq33$~meV).
We consider the ground state transition
$|2,2\rangle \rightarrow |2,1\rangle$ for a $^{87}\mathrm{Rb}$
atom with the transition frequency $f_0=\omega_{0}/2\pi=70$~kHz.
At such frequencies, the thermally-induced spin flips dominate the
spontaneous spin flips as
$\hbar\omega_0~=~4.8\,10^{-29}$~J~$(\simeq~ 0.3\mbox{ neV}) \ll~
k_BT$.
We calculate the spin matrix elements relative to that transition
through the Clebsch-Gordon coefficients and obtain for the
non-vanishing matrix elements $|\langle
i|\hat{S}_{x}|f\rangle|=|\langle i|\hat{S}_{y}|f\rangle|=1/4$.

To evaluate the Green tensor that satisfies the correct boundary
conditions at the CN surface we used the formulas obtained in
Appendix~\ref{sec:green}. In particular, the electric surface
current density creates a discontinuity in the tangential
component of the magnetic field \cite{BoundCond}. To compute the
full Green tensor, we use the method of scattering superposition
of dyadic Green tensors that are expanded into cylindrical vector
wave functions (see, e.g., \cite{GreenBook,GreenPaper}) which
somewhat differs from the approach employed in \cite{BONDAREV04}.

The lifetime increases with the atom-surface distance $y_0$ and
follows the same power law encountered in \cite{SCHEEL} for a
solid wire. According to Fig.~\ref{fig:lifetimeSF}, at an
atom-surface distance of approximately $20$~nm a lifetime of the
order of a few seconds is achievable. The spin flip lifetime can
reach one minute for distances approaching $120$~nm and exceeds
more than $100$~s for trapping distances larger than $160$~nm.
These results suggest that an atom can be held very close to a
metallic CN for sufficiently long times, and this is in line with
our expectations about spin flip occurrence and with the atom-loss
rate estimations presented in \cite{PEANO}.

\subsection{Casimir--Polder potential}

The presence of macroscopic dielectric bodies changes drastically the
structure of the vacuum electromagnetic field. One consequence is that
an atom in its ground state placed sufficiently close to a dielectric
body experiences a non-vanishing, in general attractive, dispersion
force, the Casimir--Polder (CP) force
\cite{PITAV,CP,BUHMANN-1,BUHMANN-2}. Since the CP potential adds to the
(repulsive)
trapping potential, atoms can tunnel through
the resulting potential barrier and get stuck at the nanotube
surface. The lifetime we have calculated in Sec.~\ref{sec:spinflip}
provides information about the distance at which an atom can be held
before thermally-driven spin flips occur in a given time, but the
Casimir--Polder force may play an even bigger role for small enough
distances.

The Casimir--Polder potential can be derived in lowest-order
perturbation theory within the framework of QED in dielectric
media \cite{BUHMANN-1}. If we assume that an atom is in an energy
eigenstate $|l\rangle$, then the CP potential is given by the
body-induced --- i.e. dependent on the quantity of material---
(and position-dependent) shift of the eigenvalue $\Delta E_{l}$
corresponding to this eigenstate $|l\rangle$. The CP potential can
be expressed as \cite{PITAV,BUHMANN-1}
\begin{equation}
\label{eq:vdW}
U(\mathbf{r})= \frac{\hbar \mu_0}{2 \pi} \int_{0}^{\infty} du\,u^2
\alpha_{l}^{(0)}(iu) \mathrm{Tr }[\bm{G}^{(S)}
(\mathbf{r},\mathbf{r},iu)]
\end{equation}
where $iu=\omega$ and $\alpha_{l}^{(0)}(\omega)$ is the atomic
polarizability in lowest-order perturbation theory. In
particular, for an atom in a spherically symmetric ground state, one
finds that
\begin{equation}
\label{eq:polar} \alpha_{l}^{(0)}(\omega)=\lim_{\xi \rightarrow
0}\frac{2}{3 \hbar} \sum_{k}
\frac{\omega_{kl}}{\omega_{kl}^{2}-\omega^2-i \omega \xi}
|\mathbf{d}_{lk}|^2
\end{equation}
with $\mathbf{d}_{lk}=\langle l|\hat{\mathbf{d}}|k\rangle$
representing the matrix dipole elements relative to the transition
from the atomic initial state $|l\rangle$ to the allowed states
$|k\rangle$ with frequency $\omega_{kl}\equiv(E_{k}-E_{l})/\hbar$. The
expression of the CP potential in Eq.~(\ref{eq:vdW}) is given in
terms of the scattering part $\bm{G}^{(S)}(\mathbf{r},\mathbf{r},iu)$
of the Green tensor and the frequency integral is performed along the
imaginary axis.

The Casimir--Polder potential has to be compared with the magnetic
trapping potential in order to establish the size of its effect.
\begin{figure}[ht]
\includegraphics[width=9 cm]{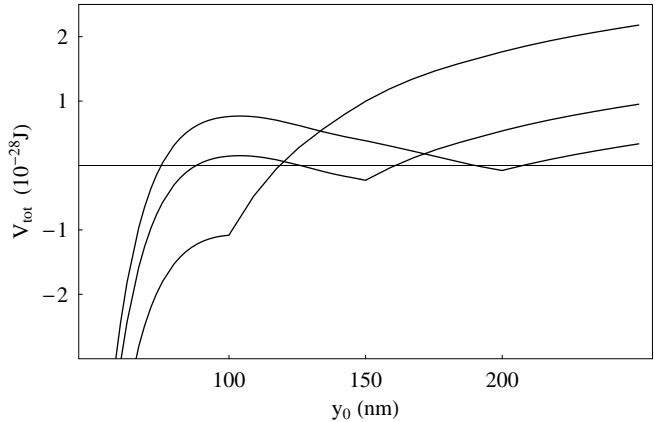}
\caption{\label{fig:Sum2} The potential $V_{\mathrm{tot}}$ is the
sum of the Casimir--Polder potential of Eq.~(\ref{eq:vdW}) and the
magnetic trapping potential of Eq.~(\ref{eq:potential}). The plot
represents $V_{\mathrm{tot}}$ for three different trapping
distances $y_{0}=100$~nm, $150$~nm, and $200$~nm, respectively. A
$(9,0)$ CN is considered with a $20\,\mu$A current, the spin-flip
transition frequency is taken to be $f_0=70$~kHz.
}
\end{figure}
In Fig.~\ref{fig:Sum2} we show $V_{\mathrm{tot}}$, the total
potential seen by the atom at three different trapping distances
$y_{0}=100$~nm, $150$~nm, and $200$~nm. $V_{\mathrm{tot}}$ is
given by the the sum of the two potentials given in
Eq.~(\ref{eq:potential}) and Eq.~(\ref{eq:vdW}). We assume a
$20\,\mu$A current is flowing through the CN as this seems to be
the largest current a single-wall CN can potentially withstand
\cite{CNcurrent} and the associated heating justifies our choice
of the temperature to be significantly higher than room
temperature. Among all the possible transitions $|l\rangle
\rightarrow |k\rangle$, we only consider the lowest electronic
transition $\mathrm{D}_2(5^2\mathrm{S}_{1/2} \rightarrow
5^2\mathrm{P}_{3/2})$ with wavelength $\lambda\simeq 780$ nm and
dipole moment $|\mathbf{d}_2|= 4.227 \, e a_0$ [$a_0$: classical
Bohr radius]. We assume that the $\mathrm{D}_{2}$ transition
represents the main contribution to the atomic polarizability
while others transitions bring about a negligible contribution to
the CP force.

In contrast to the spin-flip lifetime, temperature effects are
negligible here for two reasons. The resonant contributions
corresponding to virtual dipole absorption \cite{Ducloy06} are
suppressed because $hc/\lambda~=~2.5 \, 10^{-19}$~J~$(\simeq~
1.6\mbox{ eV})\gg~k_BT$. On the other hand, the spacing between
the Matsubara frequencies is one order of magnitude smaller than
the lowest electronic transition frequency (the relevant frequency
range over which the CP potential has to be computed) so that the
Matsubara sum can be replaced by the integral in
Eq.~(\ref{eq:vdW}).

As it is evident from Fig.~\ref{fig:Sum2}, $V_{\mathrm{tot}}$
forms a potential barrier whose height and width vary with the
trapping distance $y_0$. As mentioned previously, the addition of
the offset field $\mathbf{B}_o$ changes the bottom of the
potential well from linear to a harmonic trap which is, however,
not visible on the scale of the figure. With decreasing $y_0$ the
potential barrier becomes more and more shallow, until for
atom-surface distances smaller than the critical value of
$y_0\simeq 100$~nm the barrier effectively disappears. For
trapping distances larger than that, the total potential shows a
pronounced minimum. For example, for $y_{0}=150$ nm we estimate
the trap oscillation frequency to be $\omega_{r}\simeq 0.7$~kHz,
and the width and the height of the potential barrier to be
$68.6$~nm and $3.8\cdot 10^{-29}$~J, respectively.
\begin{figure}[ht]
\includegraphics[width=8.5cm]{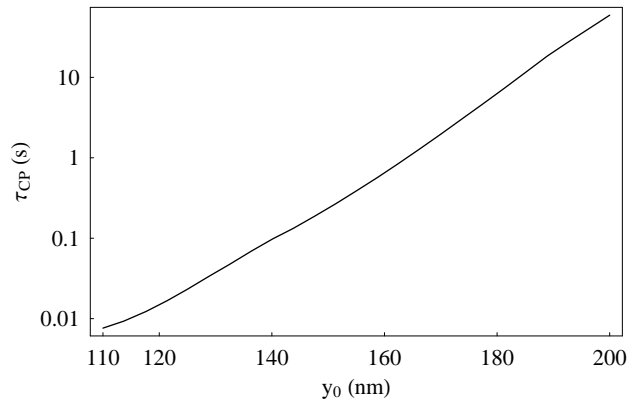} \caption{\label{fig:TunnLF}
Tunneling time $\tau_{CP}$ as a function of the trapping distance
$y_{0}$, varied between $1$ and $200$~nm.}
\end{figure}

Using the WKB approximation, we can estimate the tunneling
probability $T$ and the corresponding tunneling lifetime
$\tau_{CP}=2\pi/(T\omega_{r})$. The result is shown in
Fig.~\ref{fig:TunnLF} for a ground-state atom trapped at varying
distances $y_{0}$. From the comparison of
Fig.~\ref{fig:lifetimeSF} and Fig.~\ref{fig:TunnLF}, it is clear
that the effect of the CP force cannot be neglected. For small
enough atom-nanotube distances (and indeed for all distances shown
in the figures) the tunneling lifetime is several orders of
magnitude smaller than the spin flip lifetime. For example, at a
trapping distance $y_{0}=150$~nm we estimate $\tau_{SF}$ and
$\tau_{CP}$ to be $94.4$~s and $0.2$~s, respectively and a
tunneling lifetime of a few seconds is achievable for trapping
distances equal or bigger than $170$ nm where the spin flip
occurrence is no more a limiting factor.

\section{Conclusion}
\label{sec:conclusion}

In this paper we have investigated a novel way of miniaturizing atomic
magnetic traps by replacing solid current-carrying wires by carbon
nanotubes as the elementary building blocks. At first sight, the
advantages of using CNs are both their small diameter and the fact
that they are effectively two-dimensional structures. Hence, one
would expect from scaling arguments that traps at rather small
atom-surface distances could be realized.

We have investigated the loss mechanisms both due to thermally induced
spin flips as well as by tunneling through the Casimir--Polder
barrier. The calculations have been performed within the framework of
quantum electrodynamics in dielectric media which is valid as long as
the dielectric properties of the nanotube can be described by a
macroscopic permittivity, and if the experimental situation is such
that the atomic structure of the CN cannot be resolved and macroscopic
boundary conditions can be set.

The spin flip lifetime has been found to scale according to our
expectations. That is, this lifetime follows, as a function of the
atom-surface distance, the same power law as in the case of a solid
wire, with the result that for distances much larger than the radius of
the nanotube the expected lifetime exceeds several seconds. In
contrast, the alterations of the trapping potential by the
Casimir--Polder potential are much more severe. It appears
that the minimal feasible trapping distance is larger than
$100$~nm. The main reason explaining this result is that single-wall
nanotubes cannot sustain high enough currents (and thus cannot
generate deep enough magnetic traps) as they saturate at high electric
fields. As a potential remedy, it would be beneficial considering
multi-wall nanotubes. An increased number of carbon layers would allow
for higher current densities and consequently to a magnetic
trapping potential that would be comparable with the Casimir--Polder
potential even for smaller distances.


\acknowledgments We like to thank Lene V. Hau who, after an
initial discussion, stimulated our interest in this subject. This
work was supported by the UK Engineering and Physical Sciences
Research Council (EPSRC), partly through the UK Quantum
Information Processing Interdisciplinary Research Collaboration
(QIP IRC), and the CONQUEST and SCALA programmes of the European
commission.

\appendix

\section{Axial conductivity and dielectric permittivity}
\label{sec:epsilon} The calculation of the Casimir--Polder
potential requires the full knowledge of the frequency dependence
of the conductivity. In order to make our presentation self
contained we quote some results on calculations of the axial
conductivity that have previously been published elsewhere.

Here we briefly review the frequency dependence of the axial
surface conductivity $\sigma_{zz}(\omega)$ and of the dielectric
permittivity $\varepsilon (\omega)$ for a single-wall CN following
the presentation in \cite{TASAKI,LIN-SHUNG}. For a dielectric
medium, the linear relation between $\sigma(\omega)$ and
$\varepsilon(\omega)$ is $\varepsilon(\omega)/\varepsilon_{0}-1= i
\sigma(\omega)/(\omega \varepsilon_{0})$, where $\varepsilon_r
(\omega)=\varepsilon(\omega)/\varepsilon_{0}$ is the (complex)
relative dielectric permittivity. The Clausius-Mosotti equation
establishes the relation between the response of a medium to an
applied field, i.e. the polarization, and its dielectric constant.
Because of the cylindrical structure of CNs, their polarizability
is highly anisotropic, with the principle axis of the
polarizability tensor oriented parallel to the cylindrical axis
\cite{BENEDICT,TASAKI,DEPOL1, DEPOL2,DEPOL3}. Consequently, one is
allowed to neglect the azimuthal current \cite{LAKH}. The axial
conductivity per unit length can be expressed as
\cite{BONDAREV04,TASAKI}
\begin{eqnarray}
\label{eq:conductSI} \sigma_{zz}(\mathbf{R},\omega)&=& - \frac{i
\omega \varepsilon_{0}}{S}
\frac{\varepsilon_{r}(\mathbf{R},\omega)-1}{\rho_{T}}
\end{eqnarray}
where $\mathbf{R}=(\mathrm{R_{CN}},\phi,Z)$ is the radius vector
of an arbitrary point of the CN surface,  $S$ is the area of a
single nanotube, $\rho_T$ is the tubule density in a bundle.

 The physical properties of a CN are determined by
the way in which the graphite sheet is rolled. The winding angle
with respect to the hexagonal carbon lattice is usually described
by two integer numbers $(a,b)$. Depending on $a$ and $b$, CNs are
either semi-conducting or metallic and particularly a CN exhibits
metallic properties when $2a+b=3n$, where $n$ is again an integer
\cite{HAMADA,TASAKI,LIN-SHUNG,JIANG}. An $(a,b)$ CN has $a$
one-dimensional bands
\begin{eqnarray}\label{eq:bands}
\lefteqn{E_{\pm} (N,p)=\pm t_{0} } \nonumber
\\ && \times \sqrt{1+ 4 \cos\left
( \frac{2 \pi N}{a}- \frac{a+2b}{2 a}p \ell \right ) \cos \frac{p
\ell}{2} + 4 \cos^{2} \frac{p \ell}{2}} \nonumber
\end{eqnarray}
where $\ell$ is $\frac{3}{2}$ times the interatomic distance,
$N=0,1,\ldots,a-1$, and $\pi/\ell \leq p \leq \pi/\ell$ with $p$
the wave number. The corresponding Fermi distribution function is
$f(E)= 1/\{\exp [\beta (E-\mu)]+1\}$ with inverse temperature
$\beta$ and chemical potential $\mu$.

The main contribution to the conductivity is given by the dynamic
conductivity due to the free carrier term
$\varepsilon_{r}^{f}(\omega)$ but for high frequency regimes
another term $\varepsilon_{r}^{b}(\omega)$, arising from the
transition between the conduction and the valence bands, becomes
important such that the relative dielectric permittivity is given
by
$\varepsilon_{r}(\omega)=\varepsilon_{r}^{f}(\omega)+\varepsilon_{r}^{b}(\omega)$
\cite{TASAKI,LAKH}. The interband transition term is given by
\begin{eqnarray}
\label{eq:interbandEps} \lefteqn{ \varepsilon_{r}^{b}(\omega) = 1+
\left ( \frac{e \hbar^2}{m} \right)^{2} \frac{4 \rho c}{a \ell}
\sum_{N}
}  \nonumber \\
&& \times \int_{-\pi/l}^{\pi/l} d p \,
\frac{f(E_{+}(N,p))-f(E_{-}(N,p))}{E_{+}(N,p)-E_{-}(N,p)}
\nonumber \\
&& \times \frac{[\mathrm{Re}K_{0}(N,p)]^{2}}{(\hbar \omega)^{2}+i
\hbar^{2}\omega /\tau_{r}-[E_{+}(N,p)-E_{-}(N,p)]^{2}},
\end{eqnarray}
where  $\tau_{r}$ is a phenomenological relaxation time and
$\rho_{C}=2 a \rho_{T}= (\pi \sqrt{3})/(2 R_{\mathrm{CN}} \ell^{2})$
is the density of carbon atoms per volume \cite{JIANG}.
The Drude term is given by
\begin{equation}
\label{eq:drudeEps} \varepsilon_{r}^{f}(\omega)=-\frac{(\hbar
\omega_{pl})^{2}}{\hbar \omega (\hbar \omega+i \hbar / \tau_{r})},
\end{equation}
with $\omega_{pl}$ the plasma frequency
\begin{eqnarray}\label{eq:plasmaFr}
\omega_{pl}^{2}& =& -\left ( \frac{e \hbar}{m} \right )^{2}
\frac{2 \rho_{C}}{a \ell} \sum_{N} \int_{-\pi/\ell}^{\pi/\ell} d p
\,[\mathrm{Im} K_{0}(N,p)]^{2}
\nonumber \\
&& \times \left\{f' (E_{+}(N,p))+f' (E_{-}(N,p))\right\}.
\end{eqnarray}
The quantity $K_{0}(N,p)$ corresponds to the (dimensionless)
matrix element of the momentum operator and is given in
\cite{TASAKI}. The following parameters have been used in our
calculations: $t_{0}=4.32\times 10^{-19}$~J,
$\hbar/\tau_{r}= 4.8\times 10^{-21}$~J,
$\ell = 2.13$~{\AA}, $R_{\mathrm{CN}}=3.52$~{\AA}.

\section{Green tensor of a single-wall carbon nanotube}
\label{sec:green}

In this section we present our calculation of the dyadic Green
tensor for a single-wall CN. For a single-wall nanotube, we can
approximate the carbon layer by a boundary layer with zero
thickness. In this way, the Green tensor exhibits a discontinuity
in its first spatial derivative across the carbon layer. Due to
its cylindric symmetry, the problem can be described adopting the
cylindric basis
$\{\mathbf{e}_r,\mathbf{e}_{\varphi},\mathbf{e}_z\}$ assuming the
CN to be directed along $\mathbf{e}_z$. We use the method of
scattering superposition (see, e.g. \cite{GreenBook,GreenPaper}).
For an atom located in $\mathbf{r}'$ outside the CN, the Green
tensor can thus be written as
\begin{equation}
\label{eq:repG} \bm{G}(\mathbf{r},\mathbf{r}',\omega)
=\left\{
\begin{array}{ll}
\bm{G}_{0}(\mathbf{r},\mathbf{r}',\omega)
+\bm{G}^{(S)}_{R}(\mathbf{r},\mathbf{r}',\omega) & ,\,r>R_{CN},
\\
\bm{G}^{(S)}_{T}(\mathbf{r},\mathbf{r}',\omega) & ,\,r<R_{CN},
\end{array} \right.
\end{equation}
where $\bm{G}_{0}(\mathbf{r},\mathbf{r}',\omega)$ is the unbounded
(bulk) Green tensor representing the contribution of direct waves
from the source at $\mathbf{r}'$ to the point $\mathbf{r}$, and
the two scattering contributions
$\bm{G}^{(S)}_{R}(\mathbf{r},\mathbf{r}',\omega)$ and
$\bm{G}^{(S)}_{T}(\mathbf{r},\mathbf{r}',\omega)$ describing the
reflection and transmission of waves from/through the cylindrical
surface. In order to satisfy the homogeneous Helmholtz equation
and the radiation condition at infinity, the vacuum term and the
two scattering terms can be taken to be in the following form
\cite{GreenPaper}
\begin{eqnarray}
\label{eq:Gvacuum}
\lefteqn{
\bm{G}_{0}(\mathbf{r},\mathbf{r}',\omega)=
} \nonumber \\ &&
-\frac{\hat{\mathbf{r}} \hat{\mathbf{r}} \delta
(\mathbf{r}-\mathbf{r}')}{k^2}+\frac{i}{8 \pi}
\int^{\infty}_{-\infty} dh \sum_{n=0}^{\infty}
\frac{2-\delta_{n0}}{\eta^{2}} \nonumber \\
&&\times \left\{ \begin{array}{ll} {\mathbf{M}}^{(1)}_{^e _o n}(h)
{\mathbf{M}}'_{^e _o n}(-h)+ {\mathbf{N}}^{(1)}_{^e _o n}(h)
{\mathbf{N}}'_{^e_o n}(-h) & r>r',
\\
{\mathbf{M}}_{^e _o n}(h) {\mathbf{M}}'^{(1)}_{^e _o n}(-h)
+{\mathbf{N}}_{^e_o n}(h) {\mathbf{N}}'^{(1)}_{^e _o n}(-h) & r<r',
\end{array}\right.
\nonumber \\
\end{eqnarray}
\begin{eqnarray}
\label{eq:G11}
\lefteqn{
\bm{G}^{(S)}_{R}(\mathbf{r},\mathbf{r}',\omega)=
} \nonumber \\ &&
\frac{i}{8 \pi}
\int^{\infty}_{-\infty} dh \sum_{n=0}^{\infty}
\frac{2-\delta_{n0}}{\eta^{2}} \nonumber \\
&&\times \left\{ \left[ \mathcal{C}_{1H}
{\mathbf{M}}^{(1)}_{^e_o n}(h)+\mathcal{C}_{2H}
{\mathbf{N}}^{(1)}_{^o _e n}(h) \right]
{\mathbf{M}}'^{(1)}_{^e _o n}(-h)\right.
\nonumber \\
&& + \left. \left[ \mathcal{C}_{1V}
{\mathbf{N}}^{(1)}_{^e _o n}(h)+ \mathcal{C}_{2V}
{\mathbf{M}}^{(1)}_{^o _e n}(h) \right]
{\mathbf{N}}'^{(1)}_{^e _o n} (-h) \right\} \,,
\end{eqnarray}
\begin{eqnarray}
\label{eq:G21}
\lefteqn{
\bm{G}^{(S)}_{T}(\mathbf{r},\mathbf{r}',\omega)=
} \nonumber \\ &&
\frac{i}{8 \pi}
\int^{\infty}_{-\infty} dh \sum_{n=0}^{\infty}
\frac{2-\delta_{n0}}{\eta^{2}} \nonumber \\
&&\times \left\{ \left[ \mathcal{C}_{3H} {\mathbf{M}}_{^e _o n}(h)
+\mathcal{C}_{4H}{\mathbf{N}}_{^o _e n}(h) \right]
{\mathbf{M}}'^{(1)}_{^e _o n}(-h) \right.
\nonumber \\
&& + \left. \left[ \mathcal{C}_{3V} {\mathbf{N}}_{^e _o n}(h)+
\mathcal{C}_{4V}{\mathbf{M}}_{^o _e n}(h) \right]
{\mathbf{N}}'^{(1)}_{^e _o n} (-h) \right\} \,,
\end{eqnarray}
where $k=\omega/c$ and $\eta^2=k^{2}-h^{2}$. To enhance
readability, we have omitted the tensor product symbol $\otimes$
between the $e$ven and the $o$dd cylindrical vector wave functions
which are defined as
\begin{eqnarray}
\label{eq:NM} {\mathbf{M}}_{^e _o n}(h) &=& {\bm{\nabla}} \times
\left[ Z_{n}(\eta r) \binom{\cos}{\sin} n \phi\, e^{i h z}
\mathbf{e}_z \right],
\\
{\mathbf{N}}_{^e _o n}(h) &=& \frac{1}{k}{\bm{\nabla}} \times
{\bm{\nabla}} \times \left[ Z_{n}(\eta r) \binom{\cos}{\sin} n \phi\,
e^{i h z} \mathbf{e}_z \right ]\,.
\end{eqnarray}
The symbol $Z_{n}(x)$ has to be replaced either by the Bessel function
$J_{n}(x)$ or, if the superscript $(1)$ appears on the respective
vector wave function, by the (outgoing) Hankel function of the
first kind $H_{n}^{(1)}(x)$. The
primes in Eqs.~(\ref{eq:Gvacuum})-(\ref{eq:G21}) indicate the
cylindrical coordinates $(r',\phi',z')$. The coefficients
$\mathcal{C}_{m P}$ ($m=1$, 2, 3 and 4, and $P=H$, $V$)
need to be determined from the boundary conditions for the
electric and magnetic field components on the CN surface. The electric
field satisfies the boundary condition
\begin{equation}
\label{eq:Eboundary}
\mathbf{e}_r \times \left[
\mathbf{E}(\mathbf{r},\omega)\big|_{r=R_{\mathrm{CN}}^+}
-\mathbf{E}(\mathbf{r},\omega)\big |_{r=R_{\mathrm{CN}}^-}
\right] = 0,
\end{equation}
while the electric surface current density creates a discontinuity
in the tangential component of the magnetic field
\begin{equation}
\label{eq:Hboundary}
\mathbf{e}_r \times \left[ \mathbf{H}(\mathbf{r},\omega)
\big|_{r=R_{\mathrm{CN}}^+}
-\mathbf{H}(\mathbf{r},\omega)
\big|_{r=R_{\mathrm{CN}}^-}
\right] = \mathbf{J}(\mathbf{r},\omega) |_{r=R_{\mathrm{CN}}}.
\end{equation}
Equations~(\ref{eq:Eboundary})-(\ref{eq:Hboundary}) translate into the
respective boundary conditions for the Green tensor
\begin{equation}
\label{eq:bound1}
\mathbf{e}_r \times \left[
\bm{G}(\mathbf{r},\mathbf{r}',\omega)
\big|_{r=R_{\mathrm{CN}}^+}
-\bm{G}(\mathbf{r},\mathbf{r}',\omega)
\big|_{r=R_{\mathrm{CN}}^-}\right] = 0,
\end{equation}
\begin{eqnarray}
\label{eq:bound2}
\mathbf{e}_r \times {\bm{\nabla}} \times
\left[ \bm{G}(\mathbf{r},\mathbf{r}',\omega)
\big|_{r=R_{\mathrm{CN}}^+}
-\bm{G}(\mathbf{r},\mathbf{r}',\omega)
\big|_{r=R_{\mathrm{CN}}^-}\right] =
\nonumber \\ \hspace*{-5ex}
i \omega \mu_{0} \bm{\sigma}(\mathbf{r})\cdot
\bm{G}(\mathbf{r},\mathbf{r}',\omega)
\big|_{r=R_{\mathrm{CN}}} \,,
\end{eqnarray}
where $\bm{\sigma}(\mathbf{r})$ is the (diagonal) conductivity tensor
whose only nonzero element is $\sigma_{zz}(\mathbf{R},\omega)$.

Substituting the decomposition (\ref{eq:repG}), together with
Eqs.~(\ref{eq:Gvacuum})-(\ref{eq:G21}), into the boundary
conditions (\ref{eq:bound1}) and (\ref{eq:bound2}) leads
to two sets of four equations for each polarization $H$ and
$V$, that enable us to determine the 16 coefficients
$\mathcal{C}_{m P}$,
\begin{widetext}
\begin{eqnarray}
-\frac{\eta^{2}}{k}  H_{n}(\eta r)
\mathcal{C}_{2 H}
+ \frac{\eta^2}{k} J_{n}(\eta r) \mathcal{C}_{4 H}
&=&0,
\\
-\partial_{r} H_{n}(\eta r) \mathcal{C}_{1 H}
\pm\frac{i h n}{k r} H_{n} (\eta r)\mathcal{C}_{2 H}
+\partial_{r} J_{n}(\eta r) \mathcal{C}_{3 H}
\mp\frac{i h n}{k r}J_{n}(\eta r)
\mathcal{C}_{4 H} &=& \partial_{r} J_{n}(\eta r),
\\
-\eta^{2} H_{n}(\eta r) \mathcal{C}_{1 H}
+\eta^{2} J_{n}(\eta r)  \mathcal{C}_{3H}
&=& \eta^{2} J_{n}(\eta r),
\\
\mp\frac{i h n}{r} H_{n}(\eta r) \mathcal{C}_{1 H}
-k \partial_{r} H_{n}(\eta r)\mathcal{C}_{2H}
\pm\frac{i h n}{r} J_{n} (\eta r) \mathcal{C}_{3H}
\nonumber \\ \label{eq:bc4h}
+ \left( k \partial_{r} J_{n}(\eta r)
- i \omega \mu_{0} \sigma_{zz} \frac{\eta^{2}}{k} J_{n}(\eta r) \right)
\mathcal{C}_{4 H} &=& \pm\frac{i h n}{r} J_{n}(\eta r)
\end{eqnarray}
and
\begin{eqnarray}
-\frac{\eta^{2}}{k}  H_{n}(\eta r) \mathcal{C}_{1 V}
+ \frac{\eta^2}{k} J_{n}(\eta r) \mathcal{C}_{3V}
&=&\frac{\eta^{2}}{k} J_{n}(\eta r),
\\
\mp\frac{i h n}{k r} H_{n} (\eta r) \mathcal{C}_{1 V}
-\partial_{r} H_{n}(\eta r) \mathcal{C}_{2 V} \pm \frac{i h n}{k
r}J_{n}(\eta r) \mathcal{C}_{3 V} + \partial_{r} J_{n}(\eta
r)\mathcal{C}_{4 V} &=& \pm \frac{i h n}{k r}J_{n}(\eta r),
\\
-\eta^{2} H_{n}(\eta r) \mathcal{C}_{2V}
+ \eta^{2} J_{n}(\eta r)\mathcal{C}_{4V}
&=& 0,
\\
-k \partial_{r} H_{n}(\eta r) \mathcal{C}_{1V} \pm\frac{i h n}{r}
H_{n}(\eta r) \mathcal{C}_{2V} + \left ( k \partial_{r} J_{n}(\eta
r) - i \omega \mu_{0} \sigma_{zz} \frac{\eta^{2}}{k} J_{n}(\eta r)
\right) \mathcal{C}_{3V} \nonumber \\ \label{eq:bc4v}  \mp\frac{i
h n}{r} J_{n} (\eta r)  \mathcal{C}_{4 V} &=& k\partial_{r} J_{n}
(\eta r).
\end{eqnarray}
\end{widetext}
The appearance of the axial conductivity $\sigma_{zz}(\mathbf{R},\omega)$
in the boundary conditions (\ref{eq:bc4h}) and (\ref{eq:bc4v}) reflect
the jump condition (\ref{eq:bound2}) of the derivative of the Green tensor
at the boundary layer.

As we need to compute only the Green function in the region where the
atom is located, only $\mathcal{C}_{(1,2)P}$ need to be determined. On
using various properties of the Bessel functions such as the Wronskian
between the Bessel function $J_n(x)$ and the Hankel function
$H_n^{(1)}(x)$, $J_n(x)H_n^{(1)'}(x)-J_n'(x)H_n^{(1)}(x)=2/(\pi x)$,
we obtain that the only non-zero coefficient is
\begin{equation}
\label{eq:CV1} \mathcal{C}_{1 V}= -\frac{\pi \mu_{0} \omega
R_{\mathrm{CN}} \sigma_{zz} \eta^{2} J_{n}^{2}(\eta
R_{\mathrm{CN}})}{2 k^{2} + \pi \mu_{0} \omega R_{\mathrm{CN}}
\sigma_{zz}  \eta^{2} J_{n}(\eta R_{\mathrm{CN}}) H_{n}(\eta
R_{\mathrm{CN}})}
\end{equation}
Finally, the Green tensor for an atom located at a position $\mathbf{r}'$
outside the CN can be expressed as
\begin{eqnarray}
\label{eq:Gfinal}
\lefteqn{
\bm{G}(\mathbf{r},\mathbf{r}',\omega)=
\bm{G}_{0}(\mathbf{r},\mathbf{r}',\omega)
} \nonumber \\ &&
+\frac{i}{8 \pi} \int^{\infty}_{-\infty} dh \sum_{n=0}^{\infty}
\frac{2-\delta_{n0}}{\eta^{2}} \mathcal{C}_{1V}
{\mathbf{N}}^{(1)}_{^e _o n}(h) {\mathbf{N}}'^{(1)}_{^e _o n} (-h) \,.
\nonumber \\
\end{eqnarray}
Equation~(\ref{eq:Gfinal}), together with Eq.~(\ref{eq:Gvacuum}), is
the expression for the Green tensor used throughout this article.


\end{document}